\documentclass[10pt,conference]{IEEEtran}

\usepackage{kotex}
\usepackage{graphicx}
\usepackage{amsmath}
\usepackage{algorithm}
\usepackage{algpseudocode}
\usepackage{tabularx}
\usepackage[table]{xcolor}
\usepackage{hhline} 
\usepackage{subcaption}
\usepackage{amsfonts}
\usepackage{cite}
\usepackage{color}

\DeclareMathOperator*{\argmax}{arg\,max}
\DeclareMathOperator*{\argmin}{arg\,min}

\algnewcommand{\IIf}[1]{\State\algorithmicif\ #1\ \algorithmicthen}
\algnewcommand{\ELIIf}[1]{\State\algorithmicelse\ \algorithmicif\ #1\ \algorithmicthen}
\algnewcommand{\ElseIIf}[1]{\algorithmicelse\ #1} 
\algnewcommand{\EndIIf}{\unskip\ \algorithmicend\ \algorithmicif}

\begin{document}
\title{Robust Transmission of Punctured Text with Large Language Model-based Recovery}

\author{\IEEEauthorblockN{Sojeong Park\IEEEauthorrefmark{1},
Hyeonho Noh\IEEEauthorrefmark{2}, and
Hyun Jong Yang\IEEEauthorrefmark{2}~\IEEEmembership{Member,~IEEE}}
\IEEEauthorblockA{\IEEEauthorrefmark{1}Department of Electrical
Engineering, Pohang University of Science and Technology, Korea \\ \IEEEauthorrefmark{2} Department of Electrical and Computer Engineering, Seoul National University, Korea}
\thanks{Manuscript received December 1, 2012; revised August 26, 2015. 
Corresponding author: M. Shell (email: http://www.michaelshell.org/contact.html).}}

\maketitle

\begin{abstract}
With the recent advancements in deep learning, semantic communication which transmits only task-oriented features, has rapidly emerged. However, since feature extraction relies on learning-based models, its performance fundamentally depends on the training dataset or tasks. For practical scenarios, it is essential to design a model that demonstrates robust performance regardless of dataset or tasks. In this correspondence, we propose a novel text transmission model that selects and transmits only a few characters and recovers the missing characters at the receiver using a large language model (LLM). Additionally, we propose a novel importance character extractor (ICE), which selects transmitted characters to enhance LLM recovery performance. \textcolor{black}{Simulations demonstrate that the proposed filter selection by ICE outperforms random filter selection, which selects transmitted characters randomly.} Moreover, the proposed model exhibits robust performance across different datasets and tasks and outperforms traditional bit-based communication in low signal-to-noise ratio conditions.
\end{abstract}

\begin{IEEEkeywords}
Large language model, text transmission, data-independent, robust transmission, semantic communication.
\end{IEEEkeywords}

\section{Introduction} \label{sec:intro}

Traditional communication utilizes source coding to convert entire information into bits through lossless compression. In contrast, semantic communication, despite being a form of lossy compression, transmits only task-oriented semantic features to increase data rates while effectively maintaining task performance \cite{yang2022semantic}. In prior research, semantic features are obtained using learning-based models \cite{xie2021deep, farsad2018deep, zhang2024unified,wang2024feature, guo2023semantic}. Due to the learning-based feature extraction, semantic communication faces limitations in data robustness, similar to typical machine learning frameworks. Specifically, when a pre-trained model encounters transmitted data that was not used during the training phase, the performance degrades significantly, even in relatively simple cases of text transmission.

\begin{figure*}[t]
    \centering
    \includegraphics[width=18cm]{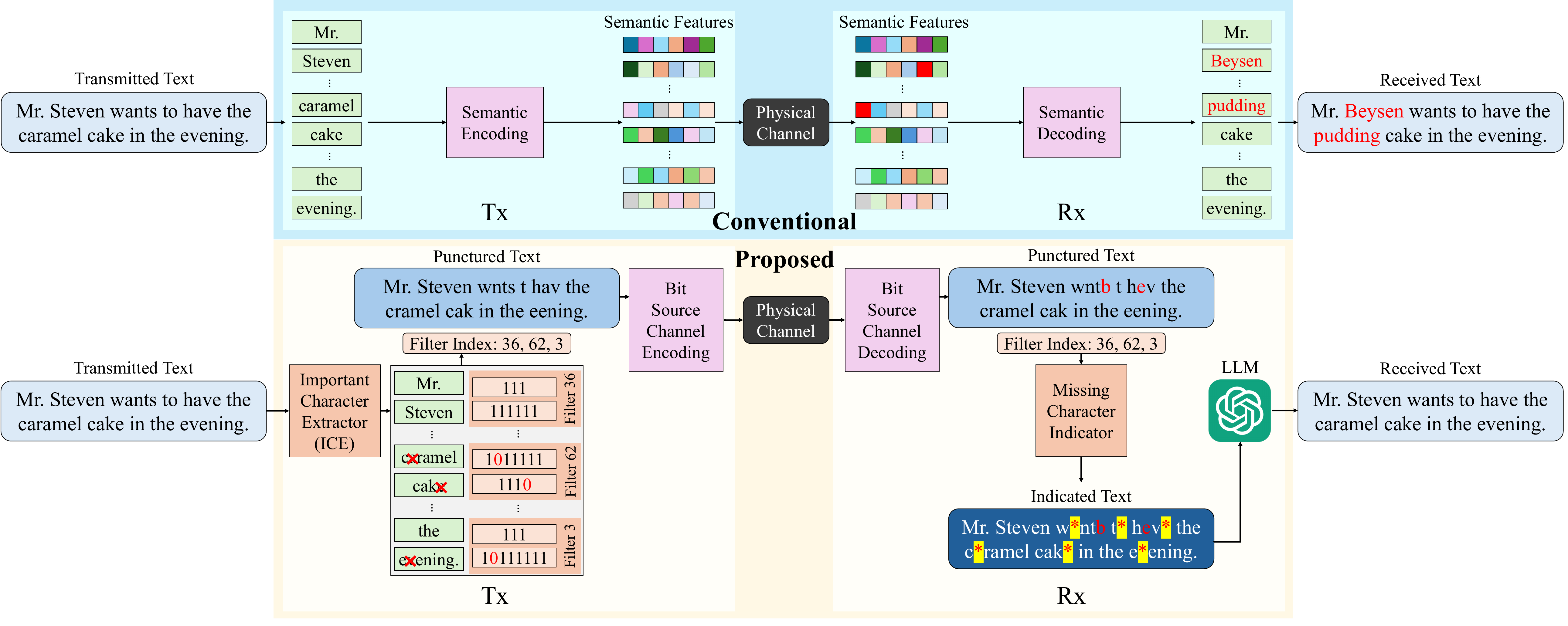}
    \vspace{-5pt}
    \caption{Comparison of the conventional semantic communication model and the proposed model.}
    \label{fig:system model}
    \vspace{-5pt}
\end{figure*}

The upper part of Fig. \ref{fig:system model} illustrates this problem: \textcolor{black}{
The conventional semantic communication model, DeepSC \cite{xie2021deep}, is trained on the Europarl dataset \cite{koehn-2005-europarl}. It utilizes transformer-based architectures to extract semantic features through word embeddings. Consequently, the distortion of features caused by channel noise leads to word-level changes for words not included in the training dataset. For instance, words `Steven' and `caramel', which are not derived from the training dataset, are inaccurately reconstructed by DeepSC. As a result, for a Q\&A task with a question like `Who wants to have the cake?', the model will produce a completely incorrect answer. Therefore, it is essential to create a robust transmission model for general tasks, regardless of the data.}

\textcolor{black}{
In this correspondence, we focus on a text transmission model that operates independently of the transmitted data. We propose a novel character-level text transmission model that transmits selected characters and reconstructs the full text using a large language model (LLM) at the receiver. If a missing character causes ambiguity between multiple possible words, the LLM may struggle to determine the correct one, potentially leading to reconstruction errors. To mitigate distortions, the proposed algorithm strategically selects characters that maximize LLM reconstruction accuracy. Furthermore, our model excludes the training process and instead leverages a pre-trained LLM that has been trained on a large volume of data to reduce dependency on specific training datasets. This extensive training enables the LLM to excel at understanding semantic information and context, even with typos or missing characters \cite{hadi2023survey}. We employ GPT-3.5 Turbo \cite{brown2020language} as a pre-trained LLM to infer the context from the received characters and reconstruct the omitted characters. The ultimate goal is to increase the transmission data rate while achieving higher task performance. Specifically, we focus on text transmission, addressing two major text-related tasks: the text reconstruction and Q\&A tasks. To this end, our contribution is as follows:}
\begin{itemize}
\item \textcolor{black}{Unlike existing transmission models that perform word-level feature extraction, the proposed model selectively transmits characters based on the proposed algorithm and reconstructs them at the receiver. To enhance recovery performance, we introduce an important character extractor (ICE) that determines which characters to be transmitted.}
\item \textcolor{black}{In conventional semantic communication, decoding relies on training process, resulting in strong dataset dependency and limiting its applicability to specific environments. In contrast, the proposed model utilizes a pre-trained LLM in the decoding process to reconstruct omitted characters, effectively reducing dataset dependency.}

\end{itemize}

\textcolor{black}{
Instead of extracting features based on word embeddings, such as transformer models, this correspondence is the first to propose a compression method through character omission combined with LLM-based recovery.} As a result, the proposed model demonstrates significantly more robust performance than conventional semantic communication models, regardless of the dataset or task. Compared to a traditional communication model, it uses fewer data symbols under the same transmission rate, allowing for more symbols for channel coding. This enables powerful error correction, resulting in better performance in low signal-to-noise ratio (SNR).

\section{System Model and Problem Formulation}
\subsection{System Model}
This correspondence considers an uplink single-input single-output system, with single antennas at the user equipment (UE) and base station (BS).
As shown in Fig. \ref{fig:system model}, ICE selects the transmitted characters, $\mathbf{p}$, from the original text, $\mathbf{t}$, and maps them into complex symbols, $\mathbf{x}$. After transmission, the received signal $\mathbf{y}$ is decoded into the estimated text, $\hat{\mathbf{t}}$, using the missing character indicator and LLM.

Specifically, the UE transmits text data, represented as $\mathbf{t} = [w_1, w_2, \dots, w_L]$, where $w_i$ denotes the $i$-th word. Each word is composed of multiple characters, thus it can be expressed as $w_i = [c^{i}_{1}, c^{i}_{2}, \dots, c^{i}_{l(w_i)}]$, where $c^{i}_{j}$ represents the $j$-th character of $w_i$, and $l(\cdot)$ indicates the length of the text. In ICE, transmitted characters are extracted using a filter selected from $M$ predefined filters. The punctured text $\mathbf{p}$, which excludes some characters, can be written as
\begin{equation}
    \mathbf{p} = I_{\epsilon}(\mathbf{t})
\end{equation}
where $I_{\epsilon}(\cdot)$ represents ICE with a compression ratio $\epsilon = l(\mathbf{t})/l(\mathbf{p})$. 
We can express the punctured text as $\mathbf{p} = [w'_1, w'_2, \dots, w'_L]$ where $w'_i$ represents the $i$-th punctured word. If the $k$-th character of $w_i$ is filtered out, then $w'_i = [c^{i}_{1}, c^{i}_{2}, \dots,c^{i}_{k-1},c^{i}_{k+1}, \dots, c^{i}_{l(w_i)}]$, and for all $i \in \{1, 2, \dots, L\}$, $l(w'_i) \le l(w_i)$ is satisfied.  For example, when `filter 62' is applied to `caramel', the result is `cramel'. The selected filter index $s$ is also transmitted, requiring only $\left\lceil \log_{2}M \right\rceil$ bits. The details of ICE will be given in section \ref{sec:ICE}. 

Next, the complex symbols $\mathbf{x} \in \mathbb{C}^{N \times 1}$ obtained through bit source and channel encoding can be represented as
\begin{equation}
    \mathbf{x} = B(\mathbf{p})
\end{equation}
where $B(\cdot)$ is the source and channel encoder and $N$ is the number of the transmitted symbols. when $\mathbf{x}$ is transmitted, the received signal $\mathbf{y} \in \mathbb{C}^{N \times 1}$ can be written as
\begin{equation}
    \mathbf{y} = h\mathbf{x} + \mathbf{n}
\end{equation}
where $h$ is a channel between the UE and BS, and $\mathbf{n}$ is an additive white Gaussian noise (AWGN), which follows $\mathcal{CN}(0, \sigma^2 \mathbf{I})$.
Using the estimated channel state information, the received signals can be recovered as $\mathbf{\hat{x}}$.

At the receiver, the estimated symbols $\mathbf{\hat{x}}$ are processed through bit source and channel decoding to obtain the recovered punctured text $\hat{\mathbf{p}} = [\hat{w'}_1, \hat{w'}_2, \dots, \hat{w'}_L]$, represented as
\begin{equation}
    \hat{\mathbf{p}} = B^{-1}(\mathbf{\mathbf{\hat{x}}})
\end{equation}
where $B^{-1}(\cdot)$ is the source and channel decoder. Then, the recovered $i$-th punctured word can be expressed as $\hat{w'}_i = [\hat{c}^{i}_{1}, \hat{c}^{i}_{2}, \dots,\hat{c}^{i}_{k-1},\hat{c}^{i}_{k+1}, \dots, \hat{c}^{i}_{l(w_i)}]$ where $\hat{c}^{i}_{j}$ represents the recovered $j$-th character of $w_i$. Additionally, the selected filter index $\hat{s}$ is recovered. Using $\hat{s}$, the missing character indicator inserts null characters into $\hat{\mathbf{p}}$, producing the indicated text $\mathbf{m}=[\hat{w}^m_1, \hat{w}^m_2, \dots, \hat{w}^m_L]$, which can be expressed as
\begin{equation}
    \mathbf{m} = M_{\hat{s}}(\hat{\mathbf{p}})
\end{equation}
where $M_{\hat{s}}(\cdot)$ is the missing character indicator with the filter index $\hat{s}$. When a null character, `*', is inserted into $\hat{w'}_i$, the text becomes $\hat{w}^m_i = [\hat{c}^{i}_{1}, \hat{c}^{i}_{2}, \dots,\hat{c}^{i}_{k-1},*,\hat{c}^{i}_{k+1}, \dots, \hat{c}^{i}_{l(w_i)}]$. For instance, `cramel' results in `c*ramel'.

Lastly, by using the LLM to replace the null characters in $\mathbf{m}$ with appropriate characters, the received text $\hat{\mathbf{t}}= [\hat{w}_1, \hat{w}_2, \dots, \hat{w}_L]$ can be obtained as follows.
\begin{equation}
    \hat{\mathbf{t}} = F(\mathbf{m})
\end{equation}
where $F(\cdot)$ refers to the LLM. The LLM recovers $\hat{w}^m_i$ as $\hat{w}_i = [\hat{c}^{i}_{1}, \hat{c}^{i}_{2}, \dots,\hat{c}^{i}_{k-1},\hat{l}^{i}_{k},\hat{c}^{i}_{k+1}, \dots, \hat{c}^{i}_{l(w_i)}]$ where $\hat{l}^{i}_{k}$ is the LLM-generated character. For example, `c*ramel' is recovered as `caramel'.

\begin{figure}[t]
    \centering
    \includegraphics[width=8.5cm]{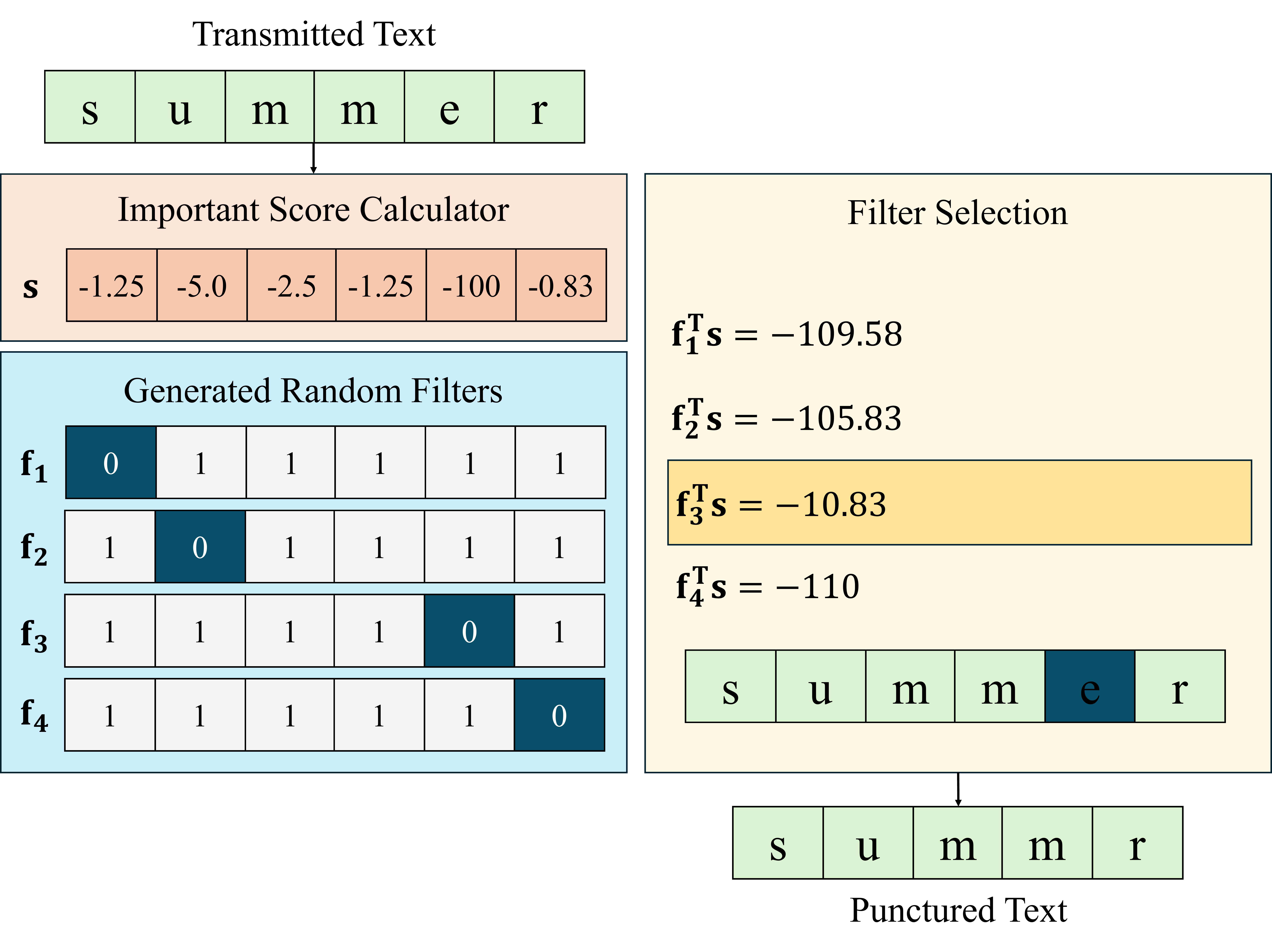}
    \vspace{-5pt}
    \caption{\textcolor{black}{Example of ICE process applied to the word `summer'.}}
    \label{fig:ICE_ex}
    \vspace{-5pt}
\end{figure}

\subsection{Problem Formulation} \label{subsec:problem formulation}
Our goal is to maximize the probability that the recovered text $\mathbf{\hat{t}}$ matches the transmitted text $\mathbf{t}$. Therefore, we define a loss function of the system as follows
\begin{equation}\label{eq:obj}
    \mathcal{L} = -\log p(\hat{\mathbf{t}} = \mathbf{t}|\mathbf{m}) = -\sum_{i=1}^{L} \log p(\hat{w}_i = w_i|\hat{w}^m_i).
\end{equation}

\textcolor{black}{
To minimize the loss, the probability of accurately reconstructing a punctured word as the transmitted word, denoted as $p(\hat{w}_i = w_i|\hat{w}^m_i)$, must be maximized. This probability is influenced by the performance of the LLM and the output of ICE, represented as $\hat{w}^m_i$. However, improving LLM performance is challenging due to computational complexity and the need for large-scale data \cite{chang2024survey}.  In contrast, ICE can be optimized by strategically selecting characters for transmission. Therefore, we focus on optimizing ICE to generate punctured text that can be effectively recovered by the LLM.}

\begin{figure}[t]
    \centering
    \includegraphics[width=8.5cm]{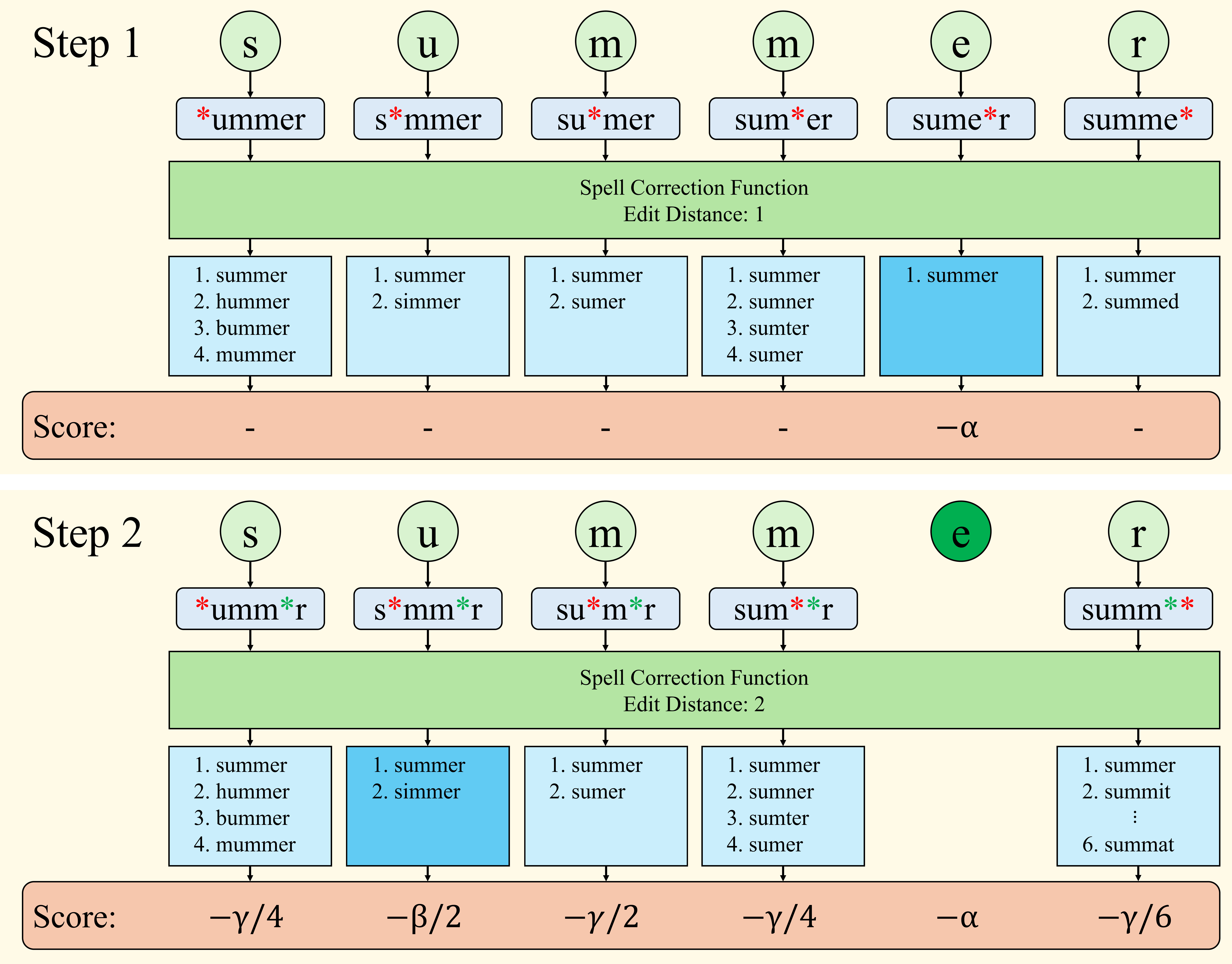}
    \vspace{-5pt}
    \caption{Example of Algorithm 1 applied to the word `summer'.}
    \label{fig:algorithm1_ex}
    \vspace{-5pt}
\end{figure}

\begin{algorithm}[t!]
    \caption{Word Character Score ($w$, $\alpha$, $\beta$, $\gamma$)}
    \label{alg:1}
\begin{algorithmic}[1]
\State \textbf{Initialize:} $\mathbf{s}_w$, $\mathbf{c}$ \Comment{First step}
\For{$i \gets 1$ to length($w$)} 
    \State $\mathcal{C} \gets \mathcal{S}(m_i, 1)$ \Comment{Possible words set}
    \IIf{$w \not\in \mathcal{C}$} $\mathbf{c}[i] \gets 0$
    \ElseIIf $\mathbf{c}[i] \gets |\mathcal{C}|$
    \EndIIf
\EndFor

\State $l_1 = \argmin_{i} \mathbf{c}[i]$

\IIf{$\mathbf{c}[l_1] = 0$} {$\mathbf{s}_w[l_1] \gets 0$}
\ELIIf{$\mathbf{c}[l_1] = 1$} {$\mathbf{s}_w[l_1] \gets -\alpha$}\\
\ElseIIf{$\mathbf{s}_w[l_1] \gets -\beta / \mathbf{c}[l_1]$}
\EndIIf

\State \textbf{Initialize:} $\mathbf{c}$ \Comment{Second step}

\For{$j \gets 1$ to length($w$)}
    \If{$j \neq l_1$}
        \State $\mathcal{C} \gets \mathcal{S}(m_{l_1,j}, 2)$ \Comment{Possible words set}
        \IIf{$w \not\in \mathcal{C}$} $\mathbf{c}[j] \gets 0$
        \ElseIIf $\mathbf{c}[j] \gets |\mathcal{C}|$
        \EndIIf
    \EndIf
\EndFor
\State $l_2 = \argmin_{j} \mathbf{c}[j]$

\IIf{$\mathbf{c}[l_2] = 0$} {$\mathbf{s}_w[l_2] \gets 0$}
\ELIIf{$\mathbf{c}[l_2] = 1$} {$\mathbf{s}_w[l_2]\gets -\alpha$}\\
\ElseIIf{$\mathbf{s}_w[l_2] \gets -\beta / \mathbf{c}[l_2]$}
\EndIIf

\For{$k \gets 1$ to length($w$)}
\If {$k \neq l_1$} and {$k \neq l_2$}
\IIf{$\mathbf{c}[k] = 0$} {$\mathbf{s}_w[k] \gets 0$}
\ElseIIf {$\mathbf{s}_w[k] \gets -\gamma / \mathbf{c}[k]$}
\EndIIf
\EndIf
\EndFor
\State \Return $\mathbf{s}_w$
\end{algorithmic}
\end{algorithm}

\begin{algorithm}[t!]
    \caption{Non-Word Character Score ($w$, $\delta$)}
    \label{alg:2}
\begin{algorithmic}[1]
\State \textbf{Initialize:} $s_w$
\State $\mathcal{C} \gets \mathcal{S}(m_a, 1)$ \Comment{Possible words set}
    \IIf{$|\mathcal{C}| = 0$} {$s_w \gets 0$}
    \ElseIIf{$s_w \gets -\delta/|\mathcal{C}|$}
    \EndIIf
\State \Return $s_w$
\end{algorithmic}
\end{algorithm}

Since $\hat{w}_i$ is generated by the LLM, calculating $p(\hat{w}_i = w_i|\hat{w}^m_i)$ requires significant time and resources, making it challenging to obtain exact values. To effectively approximate this value, we utilized a word correction method based on edit distance. Edit distance measures the minimum operations required to transform one text into another \cite{ristad1998learning}. For example, transforming `kitten' into `sitting' has an edit distance of 3. We define a spell correction function $\mathcal{S}(w, d)$, which generates a set of possible words for a corrupted input word $w$ within a maximum edit distance $d$. When $\hat{w}^m_i$ is recovered, the set of possible words can be represented as
\begin{equation}
\mathcal{S}(\hat{w}^m_i, d) = \{\hat{w}^s_1, \hat{w}^s_2, \dots, w_i, \dots, \hat{w}^s_K\}
\end{equation}
where $\hat{w}^s_l$ is a possible word of $\hat{w}^m_i$. Assume that all possible words have an equal probability to be recovered. Then $p(\hat{w}_i = w_i|\hat{w}^m_i)$ can be approximated as $1/K$. Therefore, by minimizing $K$, we can effectively reduce the loss $\mathcal{L}$. In the following section \ref{sec:ICE}, we propose an algorithm for selecting characters that minimizes loss $\mathcal{L}$ by reducing $K$.

\begin{figure}[t]
    \centering
    \begin{subfigure}[b]{0.98\columnwidth}
        \centering
        \includegraphics[width=8.5cm]{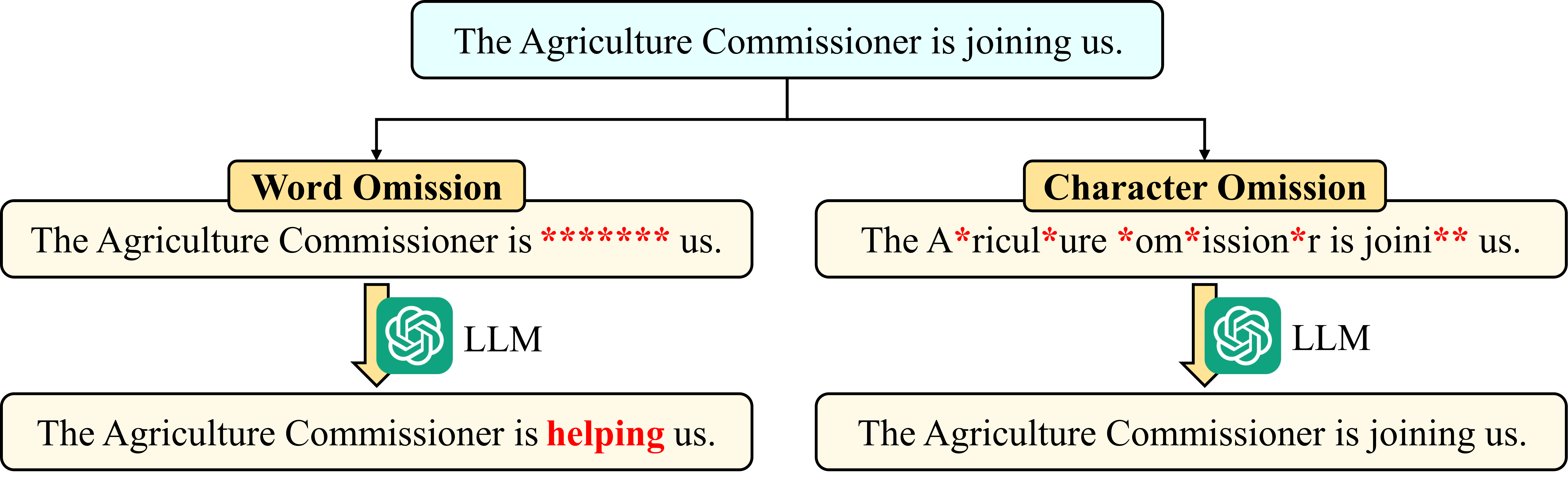}
        \caption{Example of word omission and character omission}
        \label{fig:word_character_ex}
    \end{subfigure}
    \vfill
    \begin{subfigure}[b]{0.49\columnwidth}
        \centering
        \includegraphics[width=4.5cm]{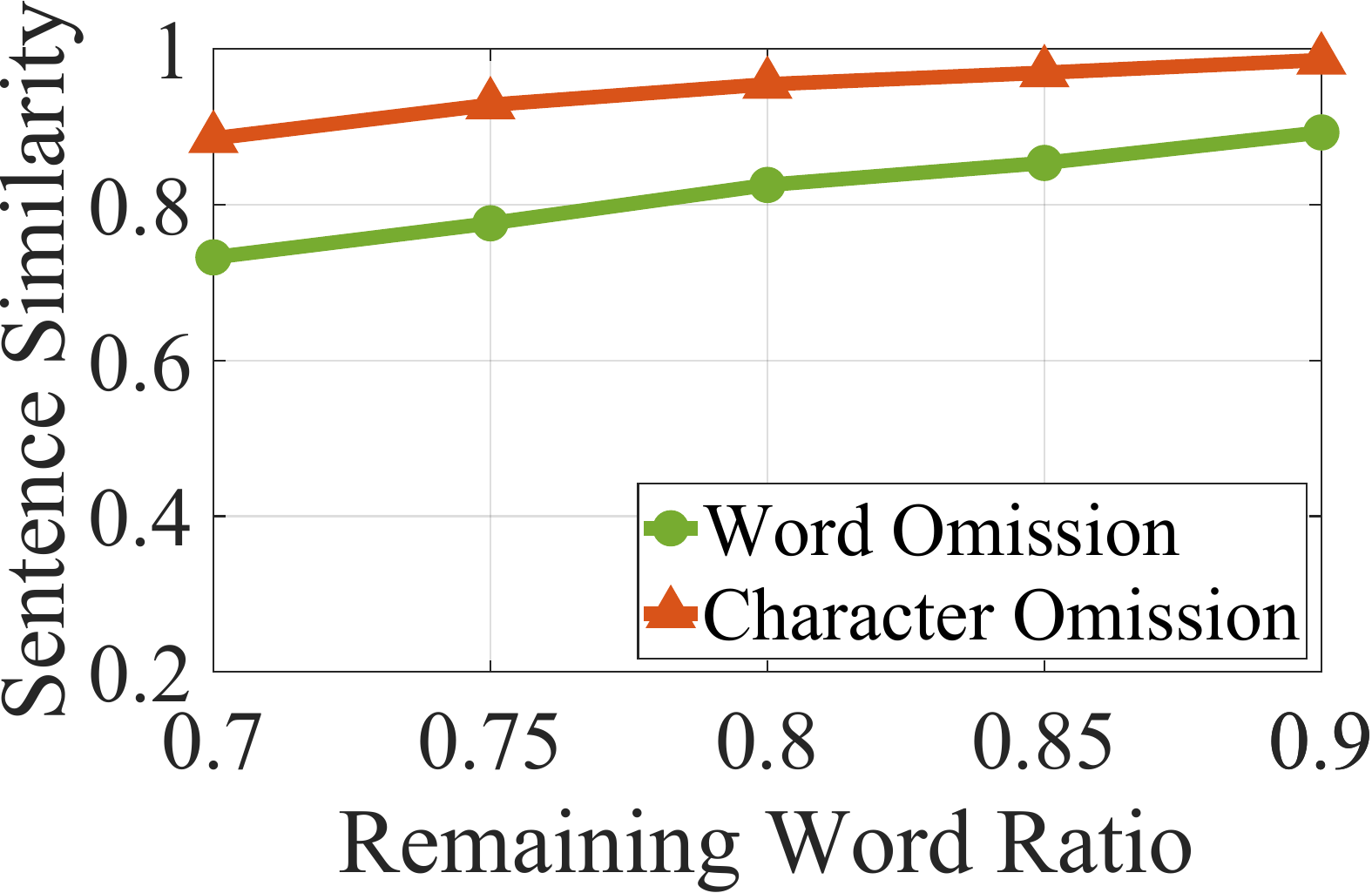} 
        \caption{Sentence Similarity versus the remaining word ratio}
        \label{fig:word_character_sim}
    \end{subfigure}
    \hfill 
    \begin{subfigure}[b]{0.49\columnwidth}
        \centering
        \includegraphics[width=4.5cm]{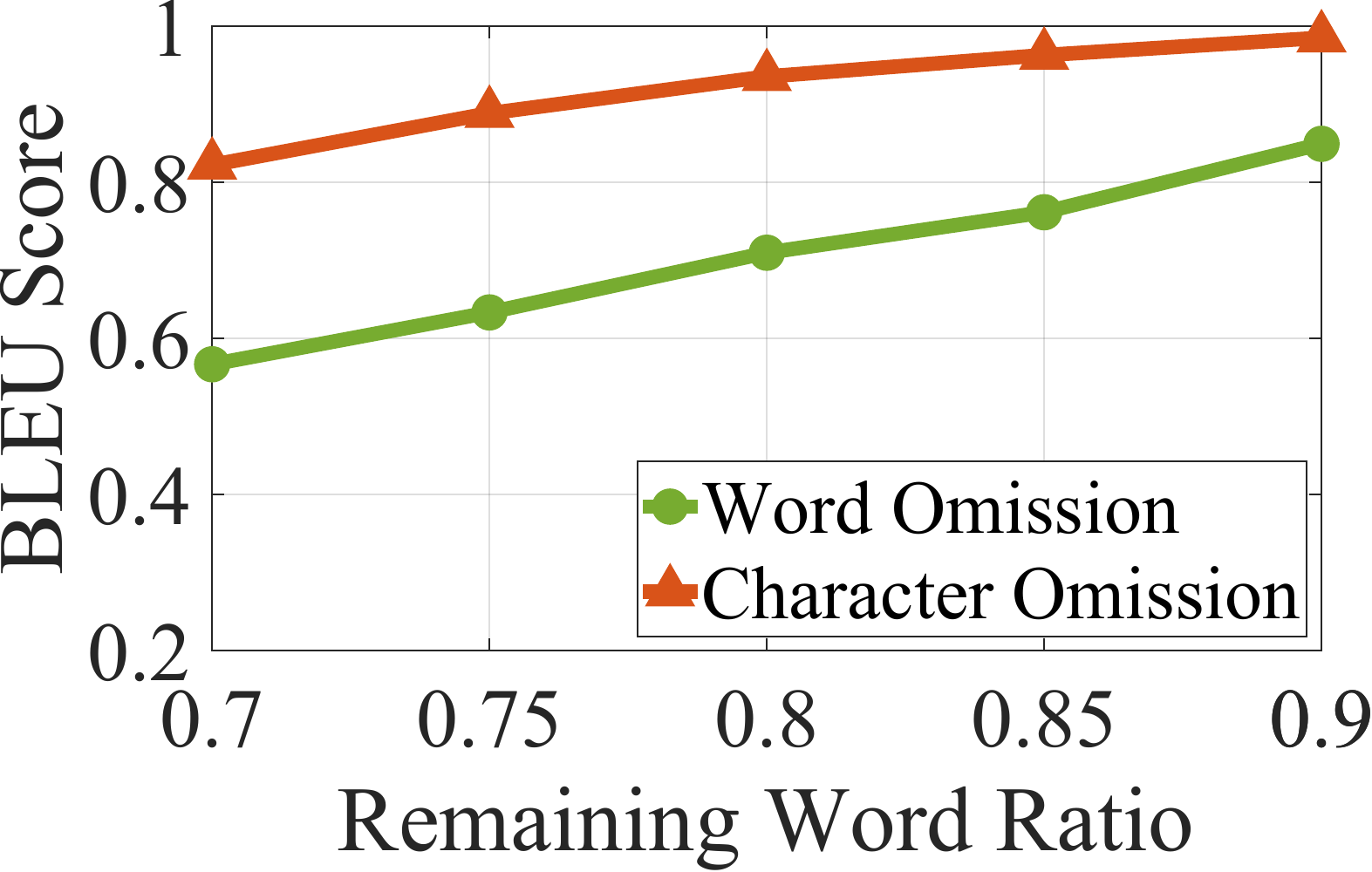}
        \caption{BLEU score versus the remaining word ratio}
        \label{fig:word_character_bleu}
    \end{subfigure}
    \caption{\textcolor{black}{Performance comparison of word omission and character omission, where the same number of characters are omitted, under different remaining word ratios in a noiseless environment.}}
    \label{fig:word_character}
\end{figure}

\section{Proposed Important Character Extractor} \label{sec:ICE}

\textcolor{black}{
ICE strategically selects characters for transmission to minimize the loss. Therefore, the input and output of ICE are the transmitted text and the punctured text, respectively.} ICE consists of a random filters generator, filter selection, and an importance score calculator, which assigns a score to each character in the text. The transmitted characters are determined based on this score. \textcolor{black}{Fig. \ref{fig:ICE_ex} illustrates the process of ICE for the word `summer'.} Omitting characters with lower scores results in a lower $K$.

\subsection{Random Filters Generator} \label{subsec:random filter}
\textcolor{black}{
In ICE, a filtering method is proposed to determine which characters are selected for transmission. The filter extracts the transmitted characters from the original text. A filter consists of zeros and ones, where zeros represent omitted characters and ones denote transmitted characters. When the length of the filter is $L_f$ and the compression ratio is $\epsilon$, the filter consists of zeros for a proportion of $(1 - 1/\epsilon)$ and ones for a proportion of $1/\epsilon$ of its total length. The random filter generator produces $M$ filters, and ICE selects the one that minimizes the loss.}

\subsection{Filter Selection} \label{subsec:filter selection}

\textcolor{black}{
In this step, the filter that maximizes the importance score of the transmitted characters is selected to optimize LLM recovery. This ensures that the important characters for LLM recovery are selected and transmitted.} The importance score is calculated by the proposed algorithm, which will be explained in following section \ref{subsec:calculator}.

The filter that maximizes the inner product with the importance score vector, $\mathbf{s} \in \mathbb{R}^{L_f \times 1}$, is selected. Let the $M$ filters generated by the random filters generator be $\mathbf{f}_1, \mathbf{f}_2, ..., \mathbf{f}_M\in\mathbb{R}^{L_f \times 1}$. Then the selected filter can be denoted as
\begin{equation} \label{eq:filter}
    \mathbf{f}_s = \argmax_{\mathbf{f} \in \{\mathbf{f}_1, \mathbf{f}_2, \ldots, \mathbf{f}_M\}} \mathbf{f}^\mathrm{T}\mathbf{s}.
\end{equation}
The selected filter is used to create the punctured text $\mathbf{p}$. \textcolor{black}{Since $\mathbf{f}_s$ is composed of $L_f/\epsilon$ ones, the length of $\mathbf{p}$ is also $L_f/\epsilon$.} This punctured text $\mathbf{p}$ and the filter index $s$ are transmitted to the receiver.

\begin{figure}[t]
    \centering
    \begin{subfigure}[b]{0.8\columnwidth}
        \centering
        \includegraphics[width=8cm]{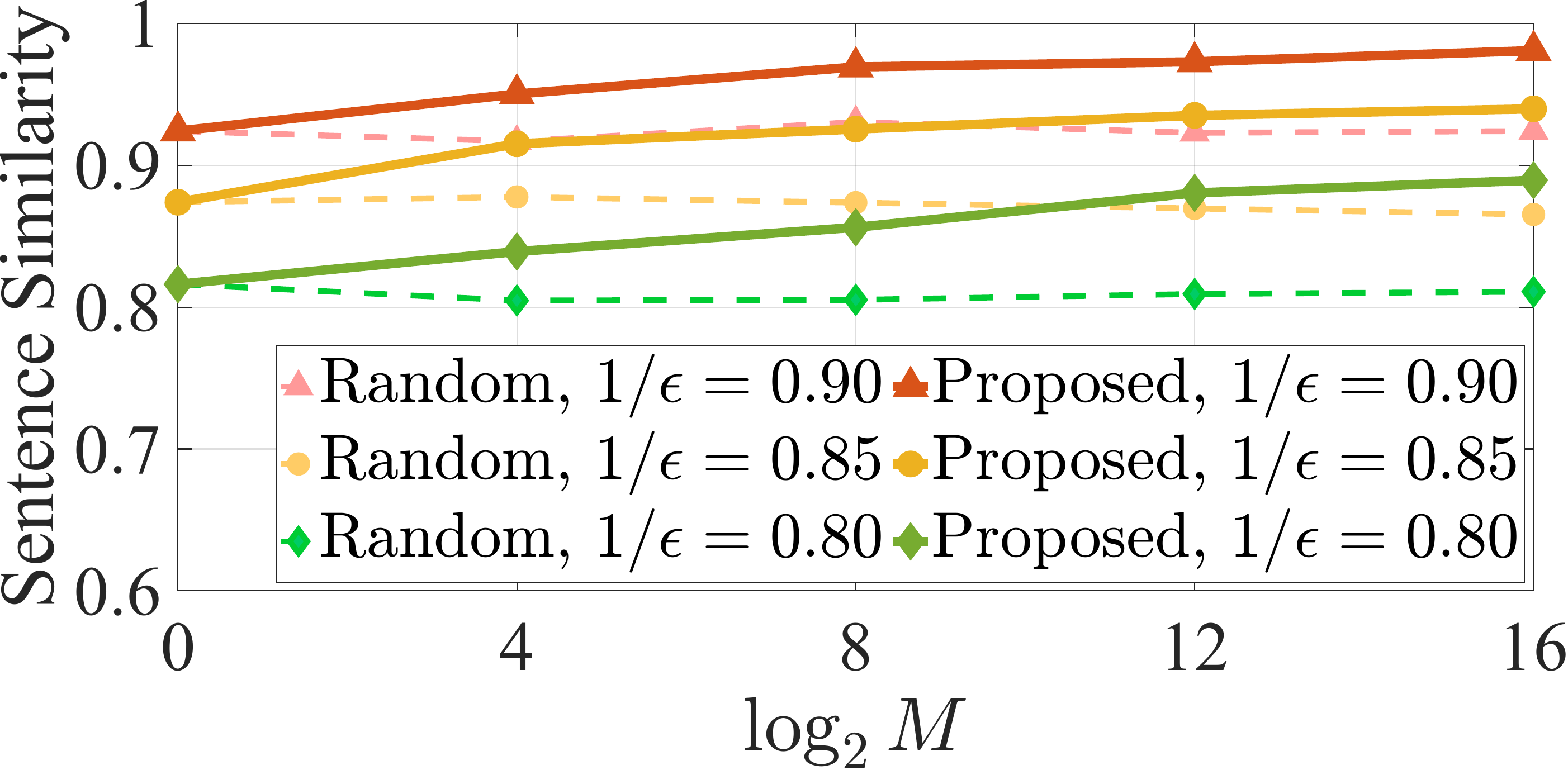} 
        \caption{Sentence Similarity versus the number of the filters}
        \label{fig:similarity_M}
    \end{subfigure}
    \vfill 
    \begin{subfigure}[b]{0.8\columnwidth}
        \centering
        \includegraphics[width=8cm]{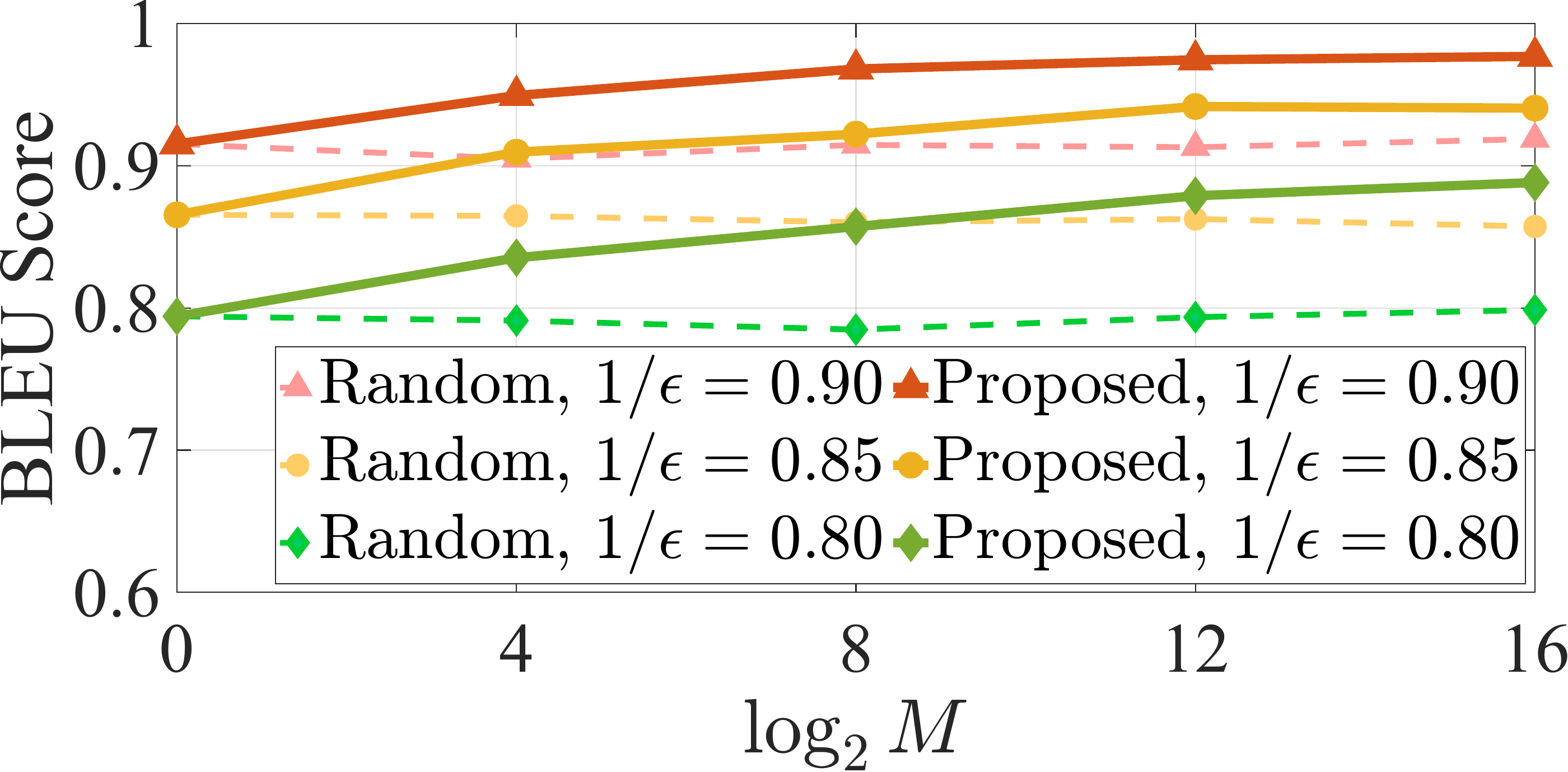}
        \caption{BLEU score versus the number of the filters}
        \label{fig:bleu_M}
    \end{subfigure}

    \caption{\textcolor{black}{Performance comparison of the proposed filter selection (solid lines) and random filter selection (dashed lines) under different compression ratios, $\epsilon$, in a noiseless environment, where $M$ denotes the number of filters.}}
    \label{fig:filter_selection}
\end{figure}

\begin{figure*}[t]
    \centering
    \begin{subfigure}[b]{0.32\textwidth}
        \centering
        \includegraphics[width=\linewidth]{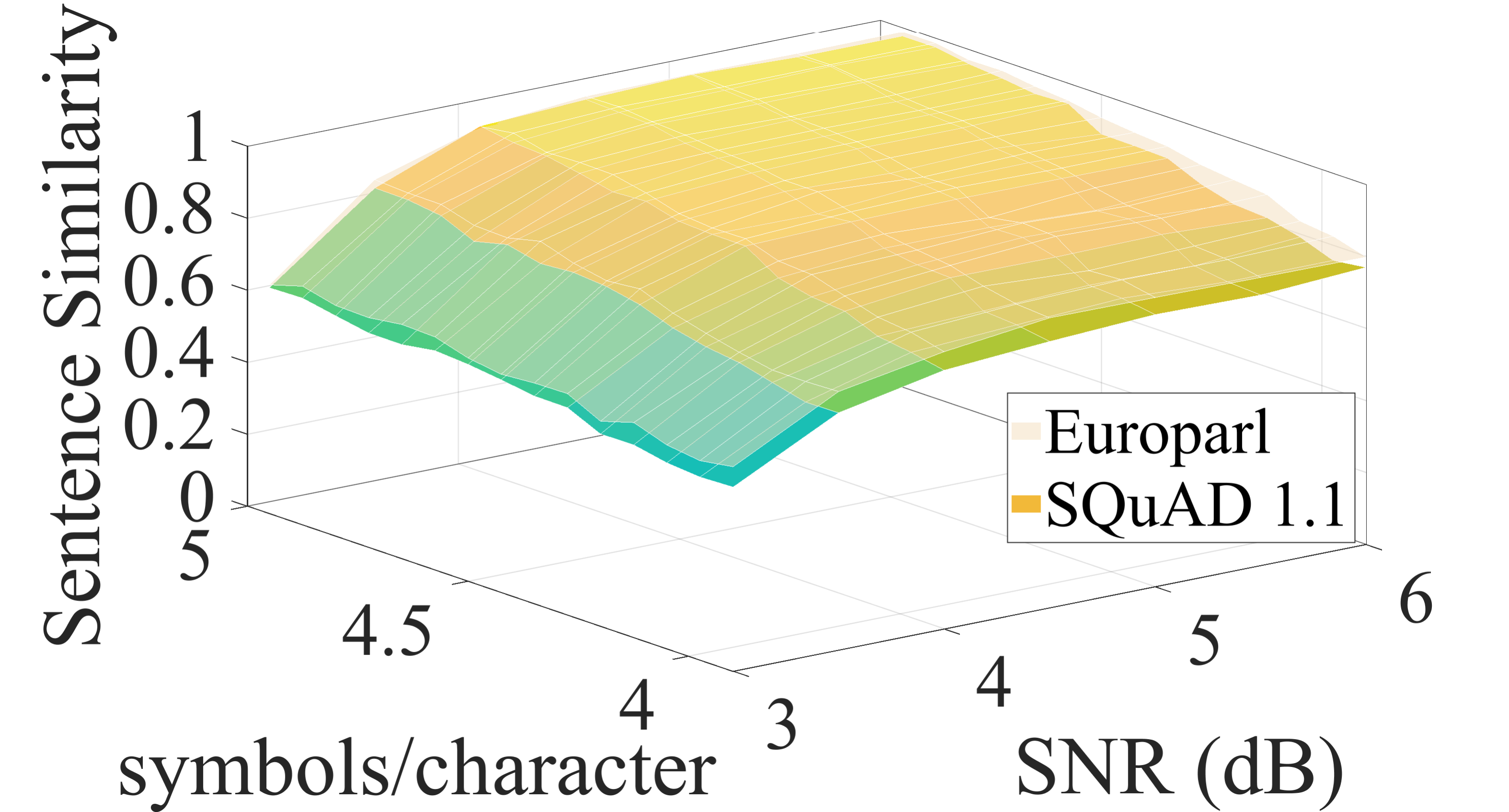} 
        \caption{Proposed model}
        \label{fig:proposed_sim}
    \end{subfigure}
    \hfill 
    \begin{subfigure}[b]{0.32\textwidth}
        \centering
        \includegraphics[width=\linewidth]{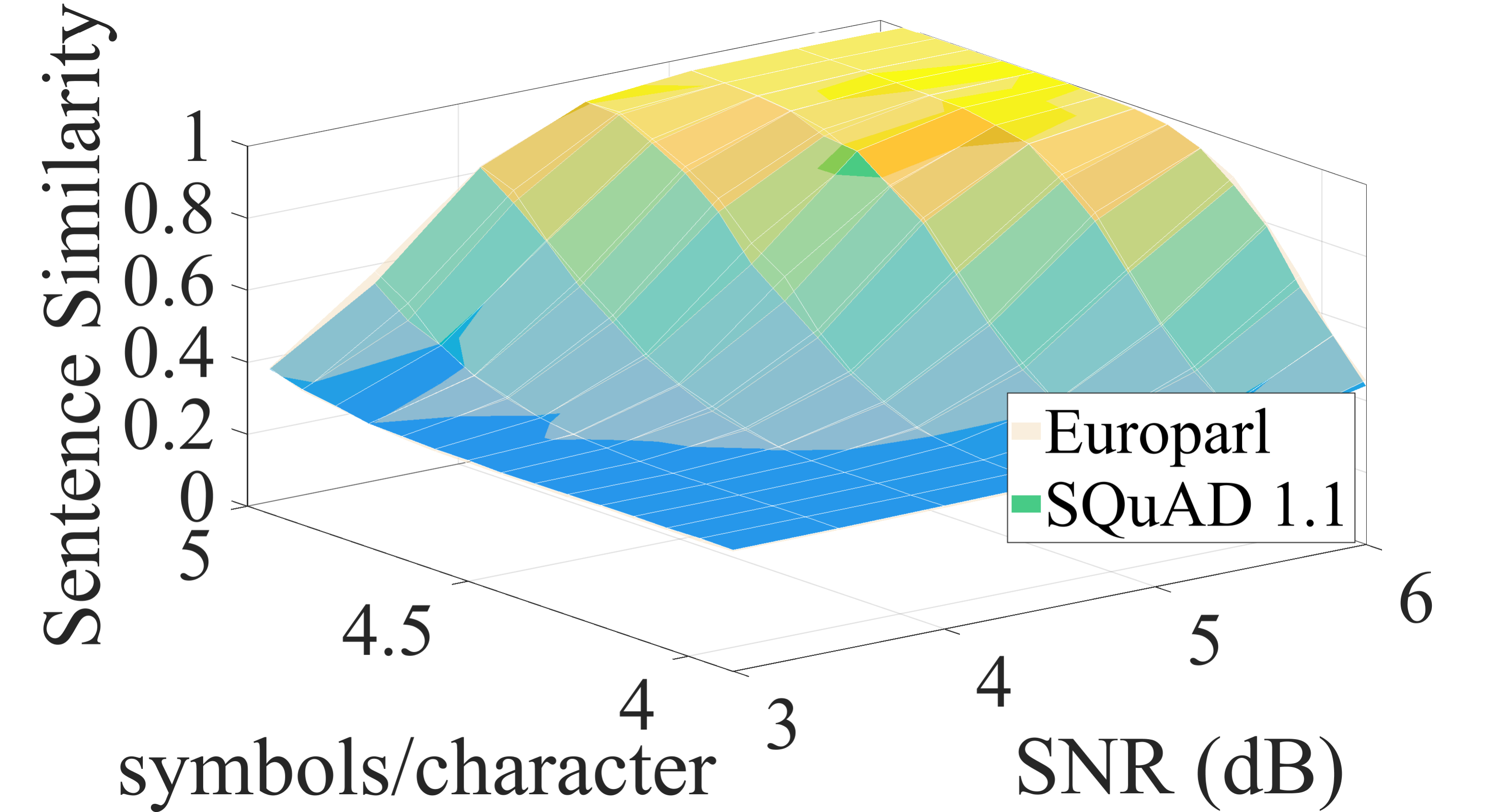}
        \caption{Traditional bit communication model}
        \label{fig:conventional_sim}
    \end{subfigure}
    \hfill
    \begin{subfigure}[b]{0.32\textwidth}
        \centering
        \includegraphics[width=\linewidth]{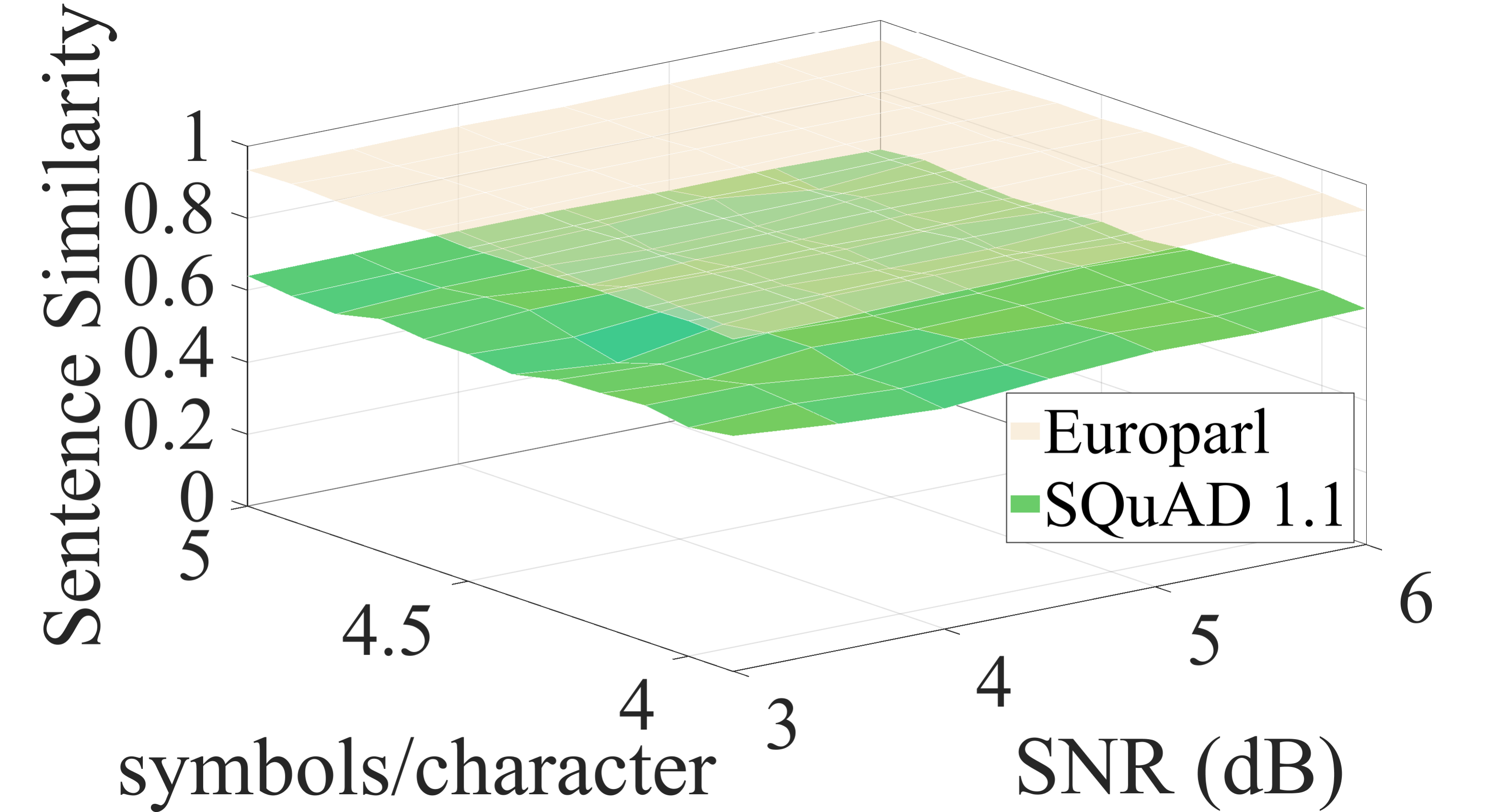} 
        \caption{DeepSC}
        \label{fig:deepsc_sim}
    \end{subfigure}

    \caption{Performance comparison of sentence similarity across the Europarl and SQuAD 1.1 datasets for the proposed model, traditional bit communication model, and DeepSC, with respect to SNR and symbols per character in an AWGN environment.}
    \label{fig:3d_sim}
\end{figure*}

\begin{figure}[t!]
    \centering
    \includegraphics[width=8.5cm]{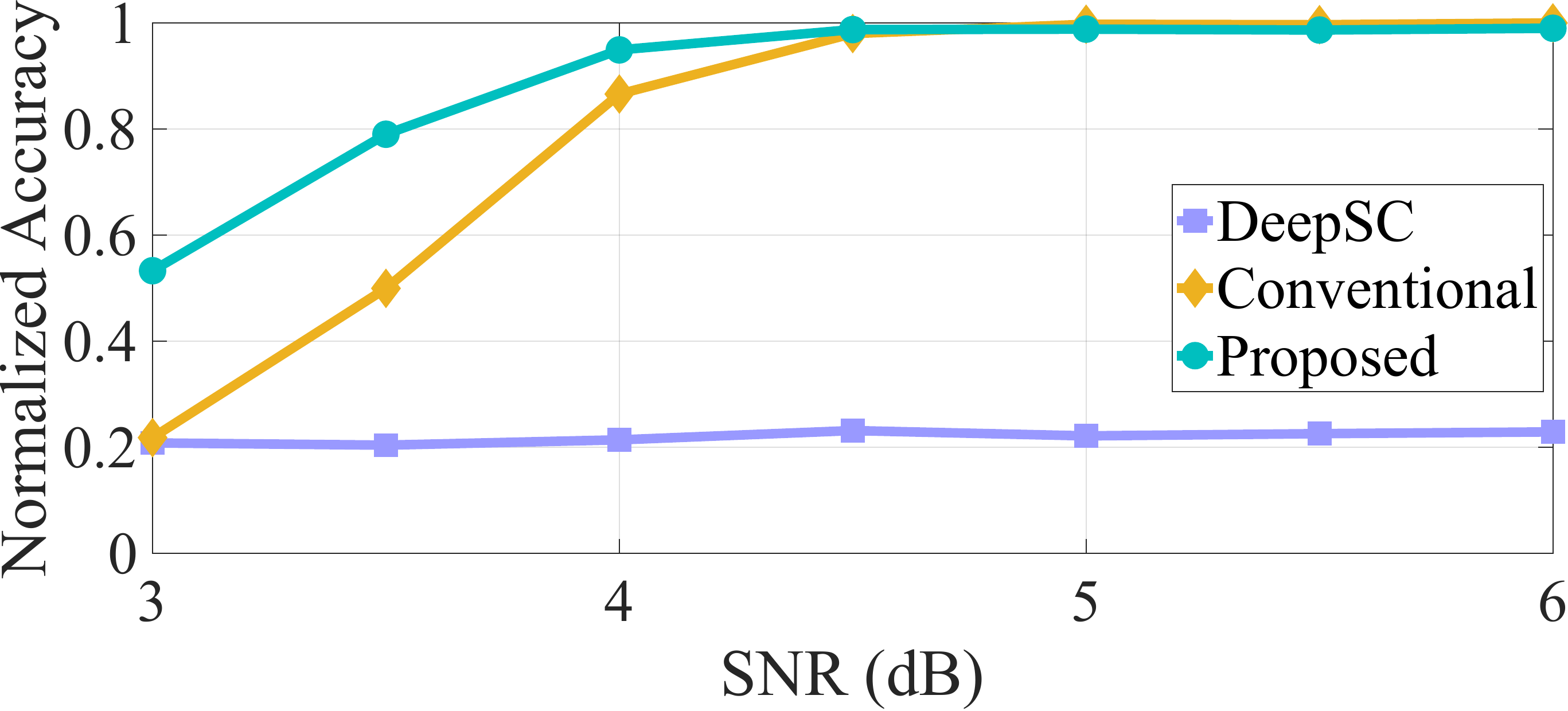}
    \vspace{-5pt}
    \caption{Normalized accuracy of the Q\&A on SQuAD 1.1 versus SNR with a fixed number of transmitted symbols.}
    \label{fig:qa_acc}
    \vspace{-5pt}
\end{figure}

\subsection{Importance Score Calculator} \label{subsec:calculator}
Filter selection is based on the importance score, making the design of the importance score calculator crucial. This section proposes algorithms for calculating text importance scores.

We define $m_i$ as the word with the $i$-th character replaced by `*', and  and $m_{i,j}$ as the word with the $i$-th and $j$-th characters replaced by `*'. Additionally, we define $m_a$ as the word with `*' appended at the end. We also define the vector $\mathbf{c}$ to store the number of possible words, the vector $\mathbf{s}_w$ to represent the importance scores of the words, and $s_w$ to represent the importance score of non-word characters.

Algorithm \ref{alg:1} is designed to calculate the importance score of each character within a word $w$ given the parameters $\alpha$, $\beta$, and $\gamma$ ($\alpha > \beta > \gamma$). Fig. \ref{fig:algorithm1_ex} shows an example of the word `summer' for Algorithm \ref{alg:1}. In Step 1, when the edit distance of the spell correction function is 1, the lowest score of $-\alpha$ is assigned to the character `e', which has only one possible word. This implies that the word can be perfectly reconstructed even if `e' is removed. In Step 2, after removing `e', the spell correction function with an edit distance of 2 is used for the remaining characters to determine the number of possible words. The character `u', which has the fewest possible words, is assigned a score of $-\beta/2$. For the remaining characters that have not been evaluated, a score of $-\gamma$ divided by the number of possible words is assigned.

Algorithm \ref{alg:2} calculates the importance score for non-word characters. The inputs are the word $w$ preceding the character to be evaluated and the parameter $\delta$ required for the importance calculation. The algorithm determines the number of words that can be corrected when a null character is added after the given word. Based on this, it assigns a score of $-\delta$ divided by the number of possible words.

\section{Simulation Results}
In this section,  we evaluate the performance of the proposed filter selection method. Additionally, we compare the proposed model with the traditional bit communication model and the conventional semantic communication model.

\subsection{Simulation Settings}

We adopt the Europarl dataset \cite{koehn-2005-europarl} and the Q\&A dataset, SQuAD 1.1 \cite{rajpurkar2016squad}. In both the proposed model and the traditional bit communication model, we use 7-bit fixed length coding for source coding, low-density parity check for channel coding, and quadrature phase shift keying for modulation. We use GPT-3.5 Turbo \cite{brown2020language} as an LLM. Additionally, we fix the length of the filter as $L_f=40$. DeepSC \cite{xie2021deep} is employed as the conventional semantic communication model.

\subsubsection{\textcolor{black}{Character omission versus word omission}} \label{subsubsec:word setting}
\textcolor{black}{Before simulating the proposed model, we compare the performance of the recovery of LLM in character omission with word omission. Character omission is determined based on the importance score derived from the proposed algorithm, and word omission is performed by randomly omitting words from the text. We conduct a performance comparison with 500 sentences derived from the Europarl dataset.}

\subsubsection{Filter selection performance}
We conduct a performance comparison with random filter selection in a noiseless environment. In this simulation, we transmit 500 sentences derived from the context of SQuAD 1.1. The simulation is carried out by varying the number of filters $M$ and the compression ratio $\epsilon$.

\subsubsection{Robustness comparison}
We use 500 sentences each from SQuAD 1.1 and the Europarl dataset, evaluating the performance of the proposed model against traditional bit communication and DeepSC in an AWGN environment. For DeepSC, we follow the model architecture described in \cite{xie2021deep} and use models trained on each dataset. Additionally, we measure the Q\&A accuracy for 1500 questions from SQuAD 1.1.

\subsection{Performance Metrics} \label{sec:metrics}

\subsubsection{BLEU Score} 
It quantifies text similarity by comparing n-grams. It is defined as
\begin{equation}\label{eq: BLEU}
    \text{BLEU} = \mathrm{BP} \cdot \exp\left(\sum_{n=1}^N w_n \log p_n\right)
\end{equation}
where $p_n$ is the n-gram score, $w_n$ the weight for each n-gram level, and $BP$ the brevity penalty, producing a value between 0 and 1, with 1 indicating a perfect match \cite{papineni2002bleu}.

\subsubsection{Sentence Similarity}
The sentence similarity \cite{xie2021deep} measures the similarity between the transmitted text $\textbf{t}$ and the received text $\hat{\textbf{t}}$, calculated as
\begin{equation}\label{eq:similarity}
    \mathrm{Similarity}(\mathbf{t},\hat{\mathbf{t}}) = \frac{\mathbf{B}(\mathbf{t}) \mathbf{B}(\hat{\mathbf{t}})^\text{T}}{\left\| \mathbf{B}(\mathbf{t}) \right\| \left\| \mathbf{B}(\hat{\mathbf{t}}) \right\|},
\end{equation}
where $\mathbf{B}(\cdot)$ is the output from BERT \cite{devlin2018bert}. This score ranges from 0 to 1, with 1 indicating maximum similarity.

\subsubsection{Normalized Accuracy}
This metric measures Q\&A accuracy on SQuAD 1.1. It is calculated by setting the accuracy from the original context with a fine-tuned DistilBERT \cite{Sanh2019DistilBERTAD} as 1 and then normalizing the accuracy from the received context of communication models, also using a fine-tuned DistilBERT, against this baseline.

\subsection{Results and Discussion}

\subsubsection{\textcolor{black}{Character omission versus word omission}} \label{subsubsec:word results}
\textcolor{black}{Fig. \ref{fig:word_character} compares the recovery performance of character omission and word omission under different remaining word ratios. If the remaining word ratio is 0.9, it means that only 90\% of the words in the text remain. For a fair comparison, the proposed algorithm is used to omit characters in character omission, ensuring that the number of omitted characters matched those in word omission. As a result, character omission achieves better performance in sentence similarity and BLEU score. This indicates that character omission based on the proposed algorithm outperforms word omission when the same amount of characters are omitted. An example of this is shown in Fig. \ref{fig:word_character_ex}.}

\subsubsection{Filter selection performance} \label{subsubsec:filter result}
\textcolor{black}{Fig. \ref{fig:filter_selection} compares random and proposed filter selection methods by measuring sentence similarity and BLEU score for different values of the number of filters, $M$, at various compression ratios, $\epsilon$. The inverse of the compression ratio, $1/\epsilon = 0.9$, means that only 90\% of the characters in the entire text are transmitted.} The proposed filter selection outperforms random selection with the same $M$ and $\epsilon$, and performance improves as $M$ increases. This shows that removing low-importance characters outperforms removing random characters.
\textcolor{black}{Additionally, the results indicate that a larger $1/\epsilon$ corresponds to better performance. In other words, a smaller compression ratio leads to improved performance.}

\subsubsection{Robustness comparison}
Fig. \ref{fig:3d_sim} illustrates the relationship between the number of transmitted symbols and SNR versus sentence similarity under two different datasets. DeepSC outperforms other models when transmitting sentences from the Europarl, as it is specifically designed for this dataset. However, its performance drops significantly when applied to SQuAD 1.1, even after retraining. This decline is due to the shallow structure of DeepSC. It compresses a vector from transformer into a lower-dimensional space using linear layers. This compression leads to information loss, particularly with large vocabularies like SQuAD 1.1. In contrast, the proposed model shows robust performance regardless of datasets. Furthermore, the proposed model outperforms traditional bit communication in low SNR. This is because it requires fewer symbols for data, allowing more symbols for channel coding.

Fig. \ref{fig:qa_acc} compares the normalized accuracy of the Q\&A task based on SNR when the number of transmitted symbols is fixed. At 3dB SNR, both the proposed model and DeepSC demonstrate sentence similarity of approximately 0.6. However, the proposed model outperforms DeepSC. This suggests that sentence similarity is not proportional to performance in other tasks. The proposed model demonstrates robust performance not only in text reconstruction but also in Q\&A tasks.

\section{Conclusion}

In this correspondence, we have proposed a punctured text transmission model with an LLM. The model selects characters for transmission based on the proposed algorithm, with the LLM recovering at the receiver. \textcolor{black}{We demonstrated that character omission outperforms word omission in LLM recovery.} Simulations further showed that the proposed filter selection outperforms random selection. Additionally, the proposed model exhibited robust performance regardless of the dataset and task. The robust performance of the proposed model indicates its potential in practical scenarios.

\bibliographystyle{IEEEbib}
\bibliography{strings,refs}

\begin{thebibliography}{10}

\bibitem{yang2022semantic}
Wanting Yang et~al.,
\newblock ``Semantic communications for future internet: Fundamentals, applications, and challenges,''
\newblock {\em IEEE Commun. Surveys Tuts}, vol. 25, no. 1, pp. 213--250, 2023.

\bibitem{xie2021deep}
Huiqiang Xie, Zhijin Qin, Geoffrey~Ye Li, and Biing-Hwang Juang,
\newblock ``Deep learning enabled semantic communication systems,''
\newblock {\em IEEE Trans. Signal Process.}, vol. 69, pp. 2663--2675, 2021.

\bibitem{farsad2018deep}
Nariman Farsad, Milind Rao, and Andrea Goldsmith,
\newblock ``Deep learning for joint source-channel coding of text,''
\newblock in {\em 2018 IEEE Int. Conf. Acoustics, Speech Signal Process. (ICASSP)}, 2018, pp. 2326--2330.

\bibitem{zhang2024unified}
Guangyi Zhang, Qiyu Hu, Zhijin Qin, Yunlong Cai, Guanding Yu, and Xiaoming Tao,
\newblock ``A unified multi-task semantic communication system for multimodal data,''
\newblock {\em IEEE Trans. Commun.}, vol. 72, no. 7, pp. 4101--4116, 2024.

\bibitem{wang2024feature}
Yining Wang et~al.,
\newblock ``Feature importance-aware task-oriented semantic transmission and optimization,''
\newblock {\em IEEE Trans. Cog. Commun. Netw.}, vol. 10, no. 4, pp. 1175--1189, 2024.

\bibitem{guo2023semantic}
Shuaishuai Guo, Yanhu Wang, Shujing Li, and Nasir Saeed,
\newblock ``Semantic importance-aware communications using pre-trained language models,''
\newblock {\em IEEE Commun. Lett.}, vol. 27, no. 9, pp. 2328--2332, 2023.

\bibitem{koehn-2005-europarl}
Philipp Koehn,
\newblock ``{E}uroparl: A parallel corpus for statistical machine translation,''
\newblock in {\em Proc. Machine Translation Summit X: Papers}, Phuket, Thailand, Sept. 13-15 2005, pp. 79--86.

\bibitem{hadi2023survey}
Muhammad~Usman Hadi et~al.,
\newblock ``A survey on large language models: Applications, challenges, limitations, and practical usage,''
\newblock {\em Authorea Preprints}, 2023.

\bibitem{brown2020language}
Tom Brown, Benjamin Mann, Nick Ryder, Melanie Subbiah, Jared~D Kaplan, Prafulla Dhariwal, Arvind Neelakantan, Pranav Shyam, Girish Sastry, Amanda Askell, et~al.,
\newblock ``Language models are few-shot learners,''
\newblock {\em Advances in neural information processing systems}, vol. 33, pp. 1877--1901, 2020.

\bibitem{chang2024survey}
Yupeng Chang et~al.,
\newblock ``A survey on evaluation of large language models,''
\newblock {\em ACM Trans. Intelligent Systems and Technology}, vol. 15, no. 3, pp. 1--45, 2024.

\bibitem{ristad1998learning}
Eric~Sven Ristad and Peter~N Yianilos,
\newblock ``Learning string-edit distance,''
\newblock {\em IEEE Trans. Pattern Analysis and Machine Intelligence}, vol. 20, no. 5, pp. 522--532, 1998.

\bibitem{rajpurkar2016squad}
Pranav Rajpurkar, Jian Zhang, Konstantin Lopyrev, and Percy Liang,
\newblock ``{SQuAD}: 100,000+ questions for machine comprehension of text,''
\newblock {\em arXiv preprint arXiv:1606.05250}, 2016.

\bibitem{papineni2002bleu}
Kishore Papineni, Salim Roukos, Todd Ward, and Wei-Jing Zhu,
\newblock ``{BLEU}: a method for automatic evaluation of machine translation,''
\newblock in {\em Proc. Annual Meeting Assoc. Comput. Linguistics (ACL)}, 2002, pp. 311--318.

\bibitem{devlin2018bert}
Jacob Devlin, Ming-Wei Chang, Kenton Lee, and Kristina Toutanova,
\newblock ``{BERT}: Pre-training of deep bidirectional transformers for language understanding,''
\newblock {\em arXiv preprint arXiv:1810.04805}, 2018.

\bibitem{Sanh2019DistilBERTAD}
Victor Sanh, Lysandre Debut, Julien Chaumond, and Thomas Wolf,
\newblock ``Distil{BERT}, a distilled version of {BERT}: smaller, faster, cheaper and lighter,''
\newblock {\em arXiv preprint arXiv:1910.01108}, 2019.

\end{thebibliography}




\end{document}